\newcommand{\diracslash}[1]{#1\llap{/\kern2pt}}

\newcommand{\be}{\begin{equation}}
\newcommand{\ee}{\end{equation}}
\newcommand{\bea}{\begin{eqnarray}}
\newcommand{\eea}{\end{eqnarray}}
\newcommand{\ba}[1]{\begin{array}{#1}}
\newcommand{\ea}{\end{array}}

\newcommand{\bt}{\begin{tabular}}
\newcommand{\et}{\end{tabular}}

\newcommand{\beas}{\begin{eqnarray*}}
\newcommand{\eeas}{\end{eqnarray*}}

\documentclass[preprint,prd,aps,floats,nofootinbib,showpacs,floatfix]{revtex4}
\usepackage{graphicx}
\begin{document}

\title{D mesons in asymmetric nuclear matter at finite temperatures}
\author{Arvind Kumar}
\email{iitd.arvind@gmail.com}
\affiliation{Department of Physics, Indian Institute of Technology, Delhi,
Hauz Khas, New Delhi -- 110 016, India}

\author{Amruta Mishra}
\email{amruta@physics.iitd.ac.in,mishra@th.physik.uni-frankfurt.de}
\affiliation{Department of Physics, Indian Institute of Technology, Delhi,
Hauz Khas, New Delhi -- 110 016, India}

\begin{abstract}
We study the in-medium properties of $D$ and $\bar{D}$ mesons in 
isospin-asymmetric nuclear matter at finite temperatures using an 
effective chiral SU(4) model. The interactions of $D$ and $\bar{D}$ 
mesons with nucleons, scalar isoscalar meson $\sigma$, 
and scalar iso-vector meson $\delta$ are taken into consideration. 
It is found that as compared to the $\bar{D}$ mesons, the properties of 
the $D$ mesons are observed to be quite sensitive to the isospin-asymmetry 
at high densities. At finite densities, the masses of $D$ and $\bar{D}$ 
mesons  observed at finite temperatures are seen to be higher in 
comparison to the zero temperature case. The present study of the 
in-medium properties of $D$ and $\bar{D}$ mesons will be of relevance 
for the experiments in the future facility of the FAIR, GSI, where the 
compressed baryonic matter at high densities and moderate temperatures 
is expected to be produced. 
The mass modifications of $D$ and $\bar{D}$ mesons in hot nuclear medium 
can lead to decay of the charmonium states  ($\Psi^{'}, \chi_{c}, J/\Psi$) 
to $D\bar{D}$ pairs in dense hadronic matter. The isospin-asymmetric effects 
on the properties of the open charm mesons for the doublets $D=(D^{0},D^{+})$ 
and $\bar D =(\bar D^0, D^-)$ should show in observables like their 
production and collective flow in asymmetric heavy-ion collisions.
The small attractive potentials observed for the $\bar{D}$ mesons 
may also lead to formation of the $\bar{D}$ mesic nuclei. 

\end{abstract}
\maketitle

\def\bfm#1{\mbox{\boldmath $#1$}}

\section{Introduction}
The study of in-medium properties of hadrons is important for 
the understanding of strong interaction physics. The study of 
in-medium hadron properties has relevance in heavy-ion collision
experiments as well as in astro-physics. There have been also
extensive experimental efforts for the study of in-medium hadron 
properties by nuclear collision experiments. In these heavy-ion 
collision experiments, hot and dense matter is produced. By studying 
the experimental observables one can infer about how the hadron 
properties are modified in the medium. For example, the observed 
enhanced dilepton spectra \cite{ceres,helios,dls} could be a 
signature of medium modifications of the vector mesons 
\cite{Brat1,CB99,vecmass,dilepton,liko}.
Similarly the properties of the kaons and antikaons have been studied 
experimentally by KaoS collaboration and the production of kaons and 
antikaons in the heavy-ion collisions and their collective flow are 
directly related to the medium modifications of their spectral 
functions \cite{CB99,cmko,lix,Li2001,K5,K6,K4,kaosnew}. The study 
of  $D$  and $\bar{D}$ mesons properties will be of direct relevance
for the upcoming experiment at FAIR, GSI, where one expects to produce
matter at high densities and moderate temperatures \cite{gsi}.
At such high densities, the properties of the $D$ and $\bar{D}$ mesons 
produced in these experiments are expected to be modified which should
reflect in experimental observables like their production and propagation
in the hot and dense medium. The reason for an expected appreciable
modifications of the $D$ and $\bar D$ mesons is
that $D$ and $\bar{D}$ mesons contain a light quark (u,d) 
or light antiquark. This light quark or antiquark interacts 
with the nuclear medium and leads to the modifications of $D$ 
and $\bar{D}$ properties. 
The experimental signature for this can be their production ratio 
and also in-medium $J/\psi$ suppression \cite{NA501,NA50e,NA502}. 
In heavy-ion collision experiments of much higher collision energies,
for example in RHIC or LHC, it is suggested that the $J/\psi$ 
suppression is because of the formation of quark-gluon plasma 
(QGP) \cite{blaiz,satz}. However, in Ref. \cite{zhang,brat5,elena} 
it is observed that the effect of hadron absorption of $J/\psi$ 
is not negligible. Due to the reduction in the masses of $D$ and 
$\bar{D}$ mesons in the medium it is a possibility that excited 
charmonium states can decay to $D\bar{D}$ pairs \cite{brat6} 
instead of decaying to lowest charmonium state $J/\psi$. 
Actually higher charmonium states are considered as major 
source of $J/\psi$ \cite{pAdata}. Even at certain higher densities,
it can become a possibility that the $J/\psi$ itself will decay 
to $D\bar{D}$ pairs. So this can be an explanation of the observed 
$J/\psi$ suppression by NA50 collaboration at $158$ GeV/nucleon 
in the Pb-Pb collisions \cite{blaiz}. The excited states of 
charmonium also undergo mass drop in the nuclear medium \cite{leeko}.
The modifications of the in-medium masses of $D$ mesons is large
then the $J/\psi$ mass modification \cite{haya1,friman}. 
This is because $J/\psi$ is
made of heavy quarks and these interact with the nuclear medium through 
gluon condensates. The change in gluon condensates with the nuclear density 
is very small and hence leads to small modification of $J/\psi$ mass.

The in-medium modifications of $D$ and $\bar{D}$ mesons has been studied in 
various approaches. For example, in the QCD sum rule approach, it is suggested 
that the light quark or antiquark of $D (\bar D)$ mesons interacts with
the light quark condensate leading to the medium modification of the 
$D(\bar D)$ meson masses \cite{arata,qcdsum08}. The quark meson coupling 
model (QMC) has also been used to study the D-meson properties \cite{qmc}. 
In the QMC model the scalar $\sigma$ meson couples to the confined 
light quark (u,d) in the nucleon thus giving  a drop of the nucleon mass
in the medium. The drop in the mass of $D$ mesons arises due to the
interaction with the nuclear medium and the mass drop of D mesons observed
in the QMC model turns out to be similar to that calculated in the 
QCD sum rule approach. 

In the present investigation we study the properties of the $D$ and $\bar{D}$ 
mesons in the isospin-asymmetric hot nuclear matter. These modifications
arise due to their interactions with the nucleons, the non-strange scalar 
isoscalar meson $\sigma$ and the scalar isovector meson $\delta$. 
The medium modifications of the light
hadrons (nucleons and scalar mesons) are described by using a chiral $SU(3)$ 
model \cite{paper3}. The model has been used to study finite nuclei, 
the nuclear matter properties, the in-medium properties of the vector mesons 
\cite{hartree,kristof1} as well as to investigate the optical potentials 
of kaons and antikaons in nuclear matter \cite{kmeson,isoamss} and in 
hyperonic matter in \cite{isoamss2}. For the study of the properties $D$ 
mesons in isospin-asymmetric medium at finite temperatures, the chiral 
SU(3) model is generalized to $SU(4)$ flavor symmetry to obtain the 
interactions of $D$ and $\bar{D}$ mesons with the light hadrons. The 
$D$ meson properties in symmetric hot nuclear matter using the chiral 
effective model have been studied in ref. \cite{amdmeson} and for the 
case of asymmetric nuclear matter at zero temperature in \cite{amarind}. 
In a coupled channel approach  for the study of D mesons,
using a separable potential, it was shown that the resonance $\Lambda_c (2593)$
is generated dynamically in the I=0 channel \cite{ltolos} analogous to
$\Lambda (1405)$ in the coupled channel approach for the $\bar K N$ interaction
\cite{kbarn}. The approach has been generalized to study the spectral density 
of the D-mesons at finite temperatures and densities \cite{ljhs},
taking into account the modifications of the nucleons in the medium. 
The results of this investigation seem to indicate a dominant increase 
in the width of the D-meson whereas there is only a very small change
in the D-meson mass in the medium \cite{ljhs}. However, these calculations
\cite{ltolos,ljhs}, assume the interaction to be SU(3) symmetric in u,d,c quarks
and ignore channels with charmed hadrons with strangeness.
A coupled channel approach for the study of D-mesons has been developed based
on SU(4) symmetry \cite{HL} to construct the effective interaction between 
pseudoscalar mesons in a 16-plet with baryons in 20-plet representation
through exchange of vector mesons and with KSFR condition \cite{KSFR}.
This model \cite{HL} has been modified in aspects like regularization method
and has been used to study DN interactions in Ref. \cite{mizutani6}. 
This reproduces the resonance  $\Lambda_c (2593)$ in the I=0 channel and in
addition generates another resonance in the I=1 channel at around 2770 MeV.
These calculations have been generalized to finite temperatures 
\cite{mizutani8} 
accounting for the in-medium modifications of the nucleons in a Walecka type
$\sigma-\omega$ model, to study the $D$ and $\bar D$ properties \cite{MK} 
in the hot and dense hadronic matter. At the nuclear matter density and for
zero temperature, these resonances ($\Lambda_c (2593)$ and $\Sigma_c (2770)$)
are generated $45$ MeV and $40$ MeV below their free space positions. 
However at finite temperature, e.g., at $T = 100$ MeV resonance positions 
shift to $2579$ MeV and $2767$ MeV for $\Lambda_{c}$  ($I = 0$) and 
$\Sigma_{c}$ ($I = 1$) respectively. Thus at finite temperature resonances 
are seen to move closer to their free space values. This is  
because of the reduction of pauli blocking factor arising due to the 
fact that fermi surface is smeared out with temperature. For $\bar{D}$ 
mesons in coupled channel approach a small repulsive mass shift is obtained. 
This will rule out of any possibility of charmed mesic nuclei \cite{mizutani8}
suggested in the QMC model \cite{qmc}. 
But as we shall see in our investigation, we obtain a small attractive mass 
shift for $\bar{D}$ mesons which can give rise to the possibility
of the formation of charmed mesic nuclei. The study of $D$ meson 
self-energy in the nuclear matter is also helpful in understanding
the properties of the charm and the hidden charm resonances in the 
nuclear matter \cite{tolosra}. In coupled channel 
approach the charmed resonance $D_{s0}(2317)$ mainly couples to $DK$ system, 
while the $D_{0}(2400)$ couples to $D\pi$ and $D_{s}\bar{K}$. The hidden 
charm resonance couples mostly to $D\bar{D}$. Therefore any modification 
of $D$ meson properties in the nuclear medium will affect the properties 
of these resonances. 

Within the effective chiral model considered in the present investigation, 
the $D(\bar D)$ energies are modified due to a vectorial Weinberg-Tomozawa, 
scalar exchange terms ($\sigma$, $\delta$) as well as range terms
\cite{isoamss,isoamss2}.
The isospin asymmetric effects among $D^0$ and $D^+$ in the doublet,
D$\equiv (D^0,D^+)$ as well as between $\bar {D^0}$ and $D^-$ in the
doublet, $\bar D \equiv (\bar {D^0},D^-)$ arise due to the scalar-isovector
$\delta$ meson, due to asymmetric contributons in the Weinberg-Tomozawa
term, as well as in the range term \cite{isoamss}.

We organize the paper as follows. In section II, we give a brief 
introduction to the effective chiral $SU(3)$ model used to study 
the isospin asymmetric nuclear matter at finite temperatures, and
its extension to the chiral $SU(4)$ model to derive the interactions
of the charmed mesons with the light hadrons. In section III, we present 
the dispersion relations for the $D$ and $\bar{D}$ mesons to be solved 
to calculate their optical potentials in the hot and dense hadronic matter.  
Section IV contains the results and discussions and finally,
in section V, we summarize the results of present investigation 
and discuss possible outlook.

\section{The hadronic chiral $SU(3) \times SU(3)$ model }

We use a chiral $SU(3)$ model for the study of the light hadrons in the 
present investigation \cite{paper3}. The model is based on nonlinear 
realization of chiral symmetry \cite{weinberg,coleman,bardeen} and 
broken scale invariance \cite{paper3,hartree,kristof1}. 
The effective hadronic chiral Lagrangian contains the following terms
\begin{equation}
{\cal L} = {\cal L}_{kin}+\sum_{W=X,Y,V,A,u} {\cal L}_{BW} + 
{\cal L}_{vec} + {\cal L}_{0} + {\cal L}_{SB}
\end{equation}
In Eq.(1), ${\cal L}_{kin}$ is the kinetic energy term, ${\cal L}_{BW}$ 
is the baryon-meson interaction term in which the baryons-spin-0 meson 
interaction term generates the baryon masses. ${\cal L}_{vec}$  describes 
the dynamical mass generation of the vector mesons via couplings to the 
scalar mesons and contain additionally quartic self-interactions of the 
vector fields. ${\cal L}_{0}$ contains the meson-meson interaction terms 
inducing the spontaneous breaking of chiral symmerty as well as a scale 
invariance breaking logarthimic potential. ${\cal L}_{SB}$ describes the 
explicit chiral symmetry breaking. 

To study the hadron properties in the present investigation we use the 
mean  field approximation and frozen glue ball limit. The scalar gluon 
condensate of QCD is simulated by a scalar dilaton field in the present
hadronic model. Since the gluon condensate is known to have a very 
marginal dependence on the changes in the density, in the present 
investigation, we take the expectation value of the dilaton field 
to be constant \cite{paper3}. This is called the frozen glue ball 
approximation. We solve the coupled equations of motion by minimizing 
the thermodynamic potential and obtain the expectation values of the 
meson fields. At finite temperatures, the vector and scalar densities 
for the baryons are given as 
\begin{eqnarray}
\rho_{i} = \gamma_{i}\int\frac{d^{3}k}{(2\pi)^{3}} 
(f_{i}(k) - \bar{f}_{i}(k))  \nonumber  \\
\rho_{i}^{s} = \gamma_{i}\int\frac{d^{3}k}{(2\pi)^{3}} 
\frac{m_{i}^{*}}{E_{i}^{*}} (f_{i}(k) + \bar{f}_{i}(k))  
\label{dens}
\end{eqnarray}
where $f_{i}(k)$ and $\bar{f}_{i}(k))$ are the thermal distribution 
functions for the baryon and antibaryon of species $i$ \cite{isoamss}
and $\gamma_i$=2 is the spin degeneracy factor.

\section{$D$ and $\bar D$ mesons in the hot asymmetric nuclear matter}

In this section we study the $D$ and $\bar{D}$ mesons properties in 
isospin-asymmetric nuclear matter at finite temperatures. 
The medium modifications of the $D$ and $\bar D$ mesons arise due
to their interactions with the nucleons and the scalar mesons and the 
interaction Lagrangian density is given as \cite{amarind}
\begin{eqnarray}
\cal L _{DN} & = & -\frac {i}{8 f_D^2} \Big [3\Big (\bar p \gamma^\mu p
+\bar n \gamma ^\mu n \Big) 
\Big({D^0} (\partial_\mu \bar D^0) - (\partial_\mu {{D^0}}) {\bar D}^0 \Big )
+\Big(D^+ (\partial_\mu D^-) - (\partial_\mu {D^+})  D^- \Big )
\nonumber \\
& +&
\Big (\bar p \gamma^\mu p -\bar n \gamma ^\mu n \Big) 
\Big({D^0} (\partial_\mu \bar D^0) - (\partial_\mu {{D^0}}) {\bar D}^0 \Big )
- \Big( D^+ (\partial_\mu D^-) - (\partial_\mu {D^+})  D^- \Big )
\Big ]
\nonumber \\
 &+ & \frac{m_D^2}{2f_D} \Big [ 
(\sigma +\sqrt 2 \zeta_c)\big (\bar D^0 { D^0}+(D^- D^+) \big )
 +\delta \big (\bar D^0 { D^0})-(D^- D^+) \big )
\Big ] \nonumber \\
& - & \frac {1}{f_D}\Big [ 
(\sigma +\sqrt 2 \zeta_c )
\Big ((\partial _\mu {{\bar D}^0})(\partial ^\mu {D^0})
+(\partial _\mu {D^-})(\partial ^\mu {D^+}) \Big )
\nonumber \\
 & + & \delta
\Big ((\partial _\mu {{\bar D}^0})(\partial ^\mu {D^0})
-(\partial _\mu {D^-})(\partial ^\mu {D^+}) \Big )
\Big ]
\nonumber \\
&+ & \frac {d_1}{2 f_D^2}(\bar p p +\bar n n 
 )\big ( (\partial _\mu {D^-})(\partial ^\mu {D^+})
+(\partial _\mu {{\bar D}^0})(\partial ^\mu {D^0})
\big )
\nonumber \\
&+& \frac {d_2}{4 f_D^2} \Big [
(\bar p p+\bar n n))\big ( 
(\partial_\mu {\bar D}^0)(\partial^\mu {D^0})
+ (\partial_\mu D^-)(\partial^\mu D^+) \big )\nonumber \\
 &+&  (\bar p p -\bar n n) \big ( 
(\partial_\mu {\bar D}^0)(\partial^\mu {D^0})\big )
- (\partial_\mu D^-)(\partial^\mu D^+) ) 
\Big ]
\label{ldn}
\end{eqnarray}
In Eq.(\ref{ldn}), the first term is the vectorial Weinberg Tomozawa 
interaction term, obtained from the kinetic term of Eq.(1). The second 
term is obtained from 
the explicit symmetry breaking term and leads to the attractive interactions 
for both the $D$ and $\bar{D}$ mesons in the medium. The next three terms of 
above Lagrangian density ($\sim (\partial_\mu {\bar D})(\partial ^\mu D)$)
are known as the range terms. The first range term (with coefficient 
$\big (-\frac{1}{f_D}\big)$) is obtained from the kinetic energy term 
of the pseudoscalar mesons. The second and third range terms $d_{1}$ 
and $d_{2}$ are written for the $DN$ interactions in analogy with 
those written for $KN$ interactions in \cite{isoamss2}. It might be 
noted here that the interaction of the pseudoscalar mesons with the 
vector mesons, in addition to the pseudoscalar meson-nucleon vectorial 
interaction, leads to a double counting in the linear realization of 
chiral effective theories. Further, in the non-linear realization,
such an interaction does not arise in the leading or subleading order, 
but only as a higher order contribution \cite{borasoy}. Hence the 
vector meson-pseudoscalar interactions will not be taken into account
in the present investigation.

The  dispersion relations for the $D$ and $\bar{D}$ mesons are obtained 
by the Fourier transformations of equations of motion. These are given as 
\begin{equation}
-\omega^{2}+\vec{k}^{2}+m_{D}^{2}-\Pi\left(\omega,\vert\vec{k}\vert\right) 
= 0
\end{equation}
where, $m_D$ is the vacuum mass of the $D(\bar D)$ meson and
$\Pi\left(\omega,\vert\vec{k}\vert\right)$ denotes the self-energy 
of the $D\left( \bar{D} \right) $ mesons in the medium.

The self-energy $\Pi\left( \omega , \vert\vec{k}\vert\right) $ for the $D$ 
meson doublet $ \left( D^{0} , D^{+}\right) $ arising from the interaction 
of Eq.(\ref{ldn}) is given as
\begin{eqnarray}
\Pi (\omega, |\vec k|) &= & \frac {1}{4 f_D^2}\Big [3 (\rho_p +\rho_n)
\pm (\rho_p -\rho_n) 
\Big ] \omega \nonumber \\
&+&\frac {m_D^2}{2 f_D} (\sigma ' +\sqrt 2 {\zeta_c} ' \pm \delta ')
\nonumber \\ & +& \Big [- \frac {1}{f_D}
(\sigma ' +\sqrt 2 {\zeta_c} ' \pm \delta ')
+\frac {d_1}{2 f_D ^2} (\rho_s ^p +\rho_s ^n)\nonumber \\
&+&\frac {d_2}{4 f_D ^2} \Big (({\rho^s} _p +{\rho^s} _n)
\pm   ({\rho^s} _p -{\rho^s} _n) \Big ) \Big ]
(\omega ^2 - {\vec k}^2),
\label{selfd}
\end{eqnarray}

where the $\pm$ signs refer to the $D^{0}$ and $D^{+}$ mesons, 
respectively, and $\sigma^{\prime}\left( \sigma - \sigma_{0}\right) $, 
$\zeta_{c}^{\prime}\left(\zeta_{c} - \zeta_{c0}\right)$, and 
$\delta^{\prime}\left(  = \delta -\delta_{0}\right) $ are the 
fluctuations of the scalar isoscalar fields $\sigma$ and $\zeta$ and
the scalar-isoscalar field $\delta$ from their 
vacuum expectation values. The vacuum expectation value of $\delta$ 
is zero $\left(\delta_{0}=0 \right)$, since a nonzero value for it 
will break the isospin-symmetry of the vacuum. (We neglect here the 
small isospin breaking effect arising  from the mass and charge 
difference of the up and down quarks.) 
We might note here that the interaction of the scalar quark condensate 
$\zeta_{c}$ (being made up of heavy charmed quarks and antiquarks)
leads to very small modifications of the masses \cite{roeder}. So we will 
not consider the medium fluctuations of $\zeta_{c}$. 
In Eq. (\ref{selfd}),
$\rho_{p}$ and $\rho_{n}$ are the number densities of protons 
and neutrons and $\rho_{p}^{s}$ and $\rho_{n}^{s}$ are their 
scalar densities, as given by equation (\ref{dens}).
  
Similarly, for the  $\bar{D}$ meson doublet 
$\left(\bar{D}^{0},D^{-}\right)$, the self-energy is calculated as 
\begin{eqnarray}
\Pi (\omega, |\vec k|) &= & -\frac {1}{4 f_D^2}\Big [3 (\rho_p +\rho_n)
\pm (\rho_p -\rho_n) \Big ] \omega\nonumber \\
&+&\frac {m_D^2}{2 f_D} (\sigma ' +\sqrt 2 {\zeta_c} ' \pm \delta ')
\nonumber \\ & +& \Big [- \frac {1}{f_D}
(\sigma ' +\sqrt 2 {\zeta_c} ' \pm \delta ')
+\frac {d_1}{2 f_D ^2} (\rho_s ^p +\rho_s ^n
)\nonumber \\
&+&\frac {d_2}{4 f_D ^2} \Big (({\rho^s} _p +{\rho^s} _n)
\pm   ({\rho^s} _p -{\rho^s} _n) \Big ]
(\omega ^2 - {\vec k}^2),
\label{selfdbar}
\end{eqnarray}
where the $\pm$ signs refer to the $\bar{D}^{0}$ and $D^{-}$ mesons, 
respectively. The optical potentials of the $D$ and $\bar{D}$ mesons are 
obtained using the expression
\begin{equation}
U(\omega, k) = \omega(k) - \sqrt{k^{2} + m_{D}^{2}}
\end{equation}
where $m_{D}$ is the vacuum mass for the $D(\bar{D})$ meson and 
$\omega(k)$ is the momentum-dependent energy of the $D(\bar{D})$ meson.

\section{Results and Discussions}
\label{results}
In this section we present the results and discussions of our investigation 
of the in-medium properties of $D$ and $\bar{D}$ mesons in isospin asymmetric 
nuclear matter at finite temperatures. We have generalized the chiral $SU(3)$ 
model to $SU(4)$ to include the interactions of the charmed mesons.
The present calculations use the following model parameters. The values, 
$g_{\sigma N} = 10.6$ and $g_{\zeta N} = -0.47$ are determined by fitting 
vacuum baryon masses. The other parameters fitted to the asymmetric 
nuclear matter saturation properties in the mean-field approximation 
are: $g_{\omega N}$ = 13.3, $g_{\rho N}$ = 5.5, $g_{4}$ = 79.7, 
$g_{\delta N}$ = 2.5, $m_{\zeta}$ = 1024.5 MeV, $ m_{\sigma}$ = 466.5 MeV 
and $m_{\delta}$ = 899.5 MeV. The coefficients $d_{1}$ and $d_{2}$, 
calculated from the empirical values of the $KN$ scattering lengths 
for $I = 0$ and $I = 1$ channels, are $2.56/m_{K}$ and $0.73/m_{K}$, 
respectively \cite{isoamss2}. 

\begin{figure}
\includegraphics[width=16cm,height=16cm]{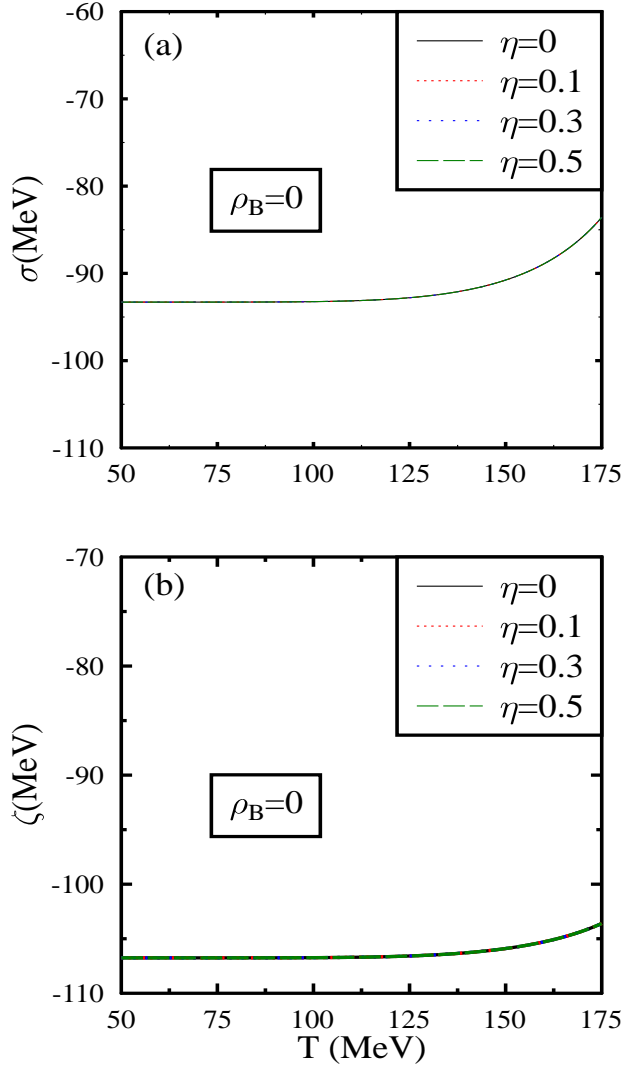}  
\caption{(Color online)
Scalar fields $\sigma$ and $\zeta$ versus temperature for baryon density 
$\rho_{B}$ = 0.
}
\label{fig1}
\end{figure}

In isospin asymmetric nuclear medium, the properties of the $D$ mesons 
($D^{+}$, $D^{0}$) and $\bar{D}$ mesons ($D^{-}$,$\bar{D}^{0}$), 
due to their interactions with the hot hadronic medium, undergo 
medium modifications. These modifications arise due to the interactions
with the nucleons (through the Weinberg-Tomozawa vectorial interaction
as well as through the range terms) and the scalar exchange terms.
The modifications of the scalar mean fields modify the masses of the
nucleons in the hot and dense hadronic medium. Before going into the 
details of how the $D$ and $\bar{D}$ mesons properties are modified 
at finite temperatures in the dense nuclear medium, let us see how 
the scalar fields are modified at finite temperatures in the nuclear 
medium. In figure \ref{fig1}, we plot the variations of $\sigma$ 
and $\zeta$ with temperature at zero baryon density.
We observe that the magnitudes of the scalar fields $\sigma$ and $\zeta$ 
decrease with increase in temperature. However, the drop in their magnitudes
with temperature is negligible upto a temperature of about 100 MeV
(that is, they remain very close to their vacuum values).
The changes in the magnitudes of the $\sigma$ and $\zeta$ 
fields are 2.5 MeV and 0.8 MeV respectively, when the temperature
changes from 100 MeV to 150 MeV, above which the drop increases.
These values change to about 10 MeV and 3 MeV for the $\sigma$
and $\zeta$ fields respectively, for a change in temperature of 
100 MeV to 175 MeV.

\begin{figure}
\includegraphics[width=16cm,height=16cm]{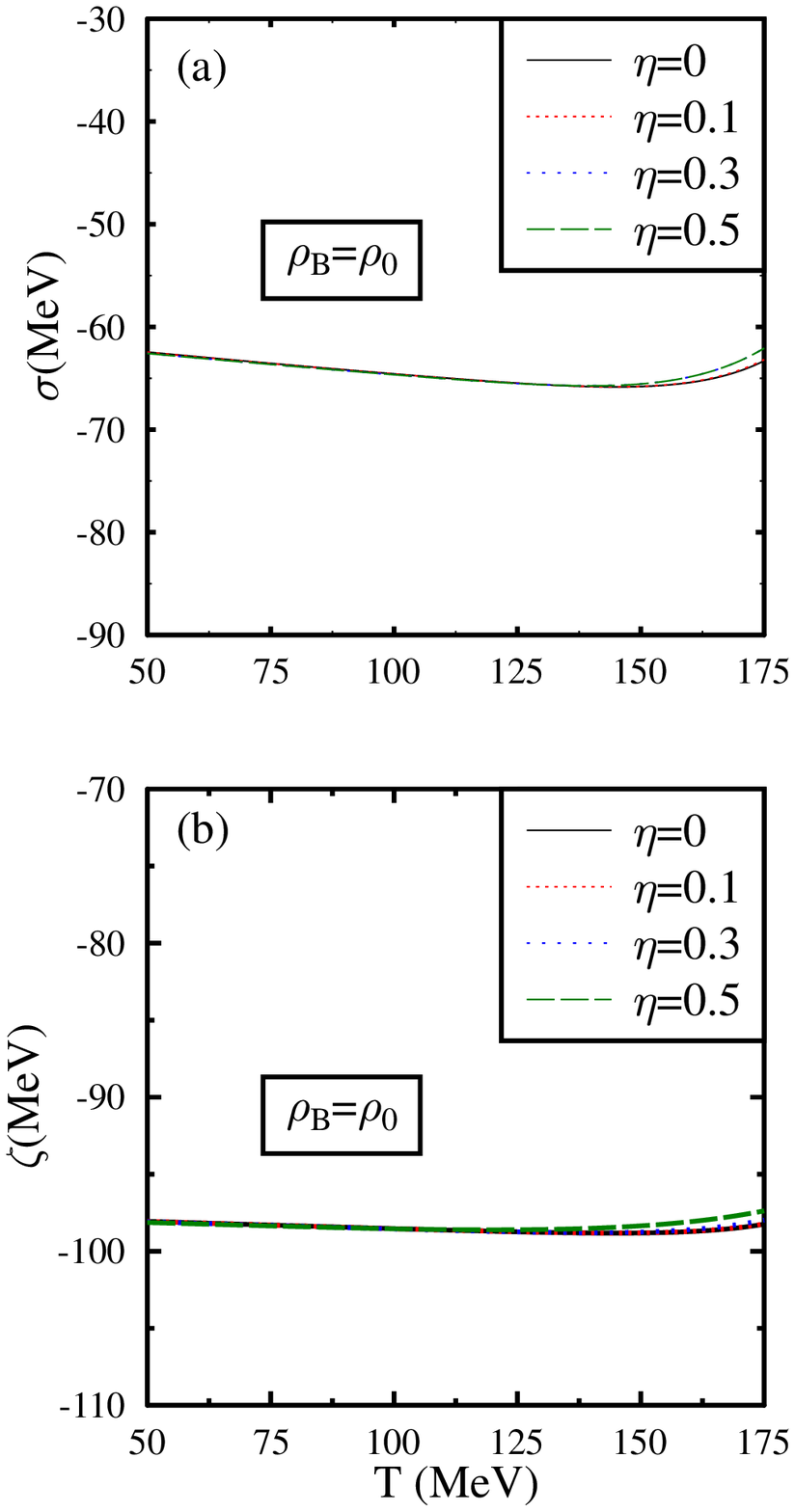} 
\caption{(Color online)
Scalar fields $\sigma$ and $\zeta$ versus temperature for density 
$\rho_{B}$ = $\rho_0$ and isospin asymmetric parameter $\eta$ as 
0, 0.1, 0.3 and 0.5.
}
\label{fig2}
\end{figure}

\begin{figure}
\includegraphics[width=16cm,height=16cm]{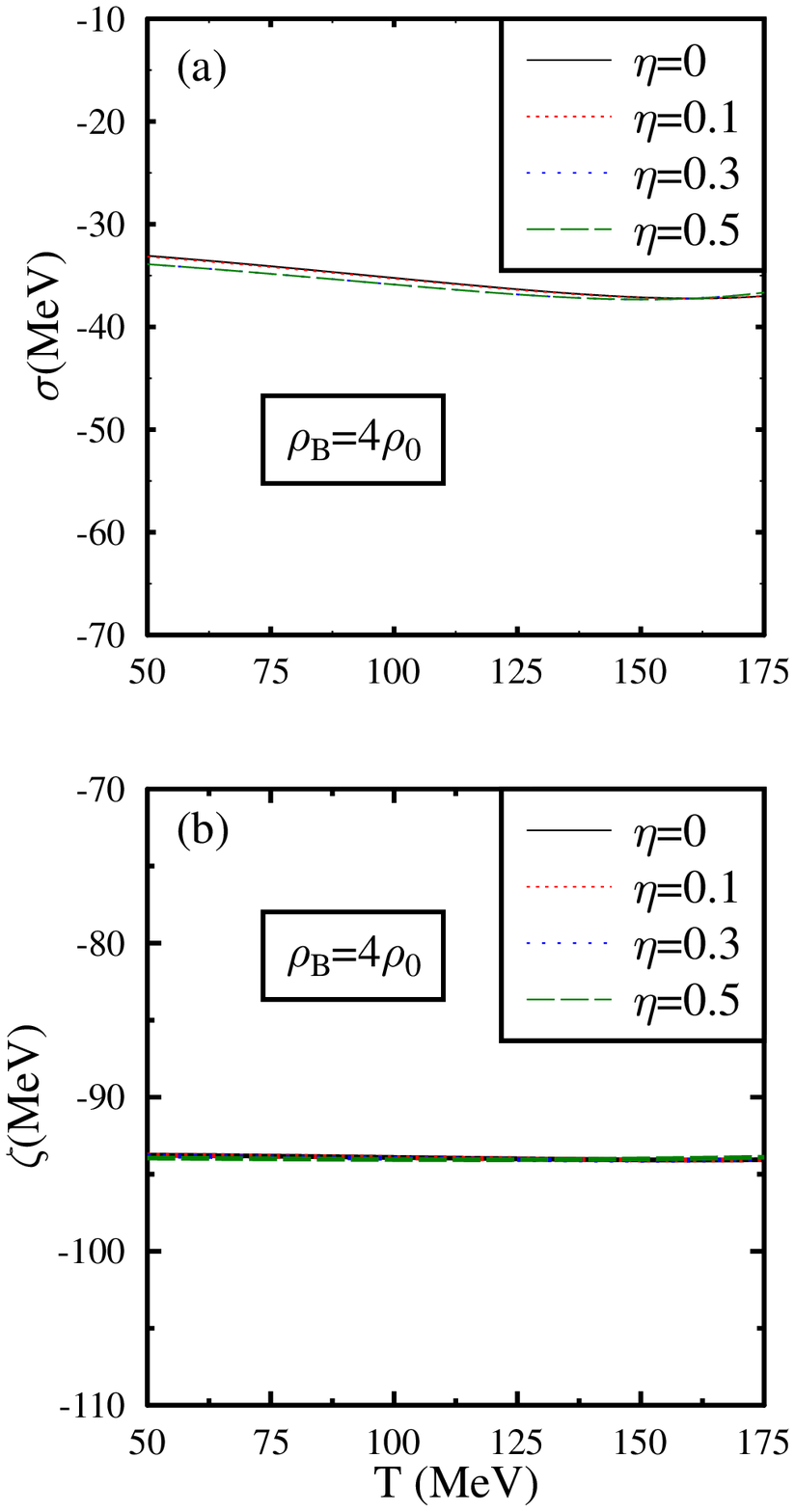} 
\caption{(Color online)
Scalar fields $\sigma$ and $\zeta$ versus temperature for density 
$\rho_{B}$ = 4 $\rho_0$ and isospin asymmetric parameter $\eta$ as 
0, 0.1, 0.3 and 0.5.
}
\label{fig3}
\end{figure}

We plot the temperature dependence of the $\sigma$ and $\zeta$ fields 
for baryon densities $\rho_B$=$\rho_0$ and $\rho_B=4\rho_0$ in figures
\ref{fig2} and \ref{fig3} respectively. At finite values of 
densities, the magnitudes of the scalar fields $\sigma$ and $\zeta$ first 
increase with increase in temperature upto a certain value after which 
they start decreasing. For example, at $\rho_{B} = \rho_{0}$, 
and for the value of the isospin asymmetry parameter, $\eta =  0$ 
the scalar fields $\sigma$ and $\zeta$ increase with 
temperature upto a temperature of about 150 MeV and 
after they both start decreasing. At $\eta =  0.5$ the value 
of temperature upto which these scalar fields increase becomes 
about 120 -- 125 MeV. For $\rho_{B} = 4\rho_{0}$ and $\eta = 0$ 
the scalar field $\sigma$ 
and $\zeta$ fields increase upto a temperature of about 160 MeV.
At $\eta = 0.5$ this value of temperature is lowered to 120 MeV 
and 135 MeV for $\sigma$ and $\zeta$ respectively. 
This observed rise in the magnitudes of $\sigma$ 
and $\zeta$ fields with temperature leads to an increase in the mass of 
nucleons with temperature at finite densities. Such a behaviour of nucleon 
mass with temperature at finite baryon densities was also observed 
within chiral $SU(3)$ model in \cite{kristof1} 
and within Walecka model by Ko and Li in \cite{liko}. 


\begin{figure}
\includegraphics[width=16cm,height=16cm]{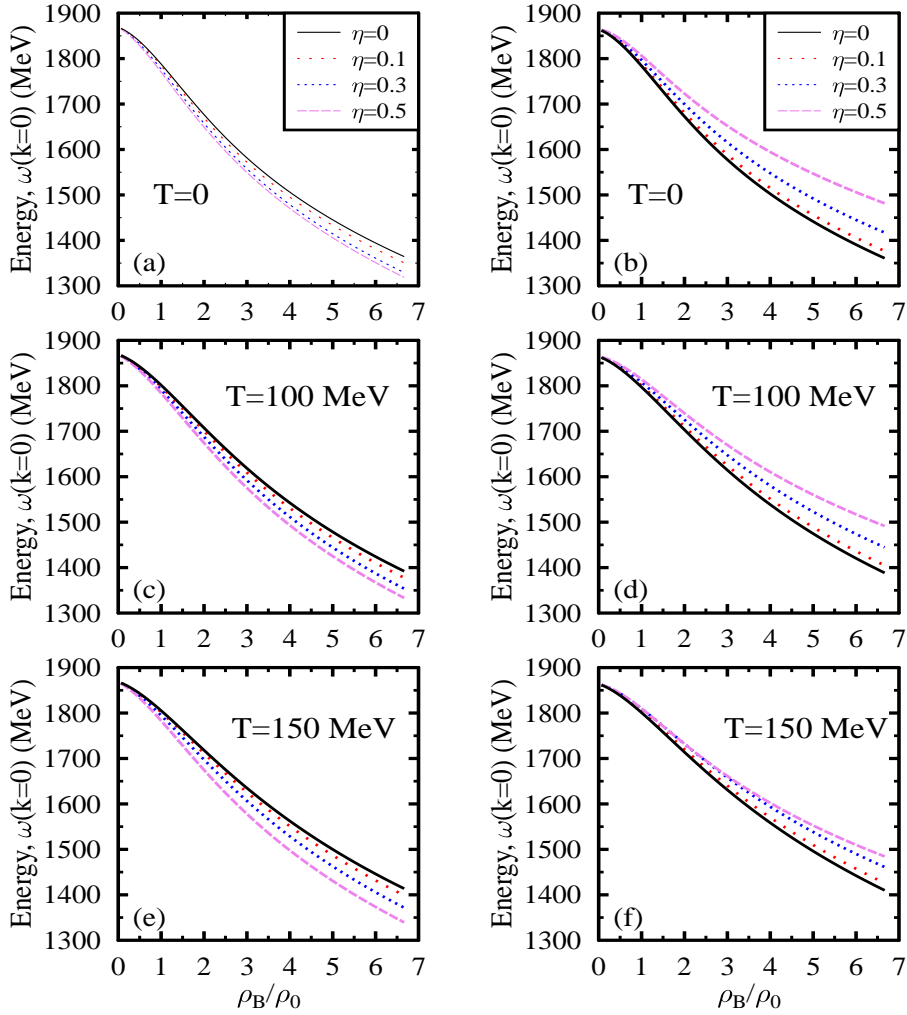} 
\caption{(Color online) The energies of $D^{+}$ meson ((a),(c) and (e)) 
and of $D^{0}$ meson ((b),(d) and (f)), at momentum $k = 0$, versus 
the baryon density (in units of nuclear saturation density), 
$\rho_{B}/\rho_{0}$, for different values of the isospin asymmetry 
parameter ($\eta = 0, 0.1, 0.3, 0.5$) and for given values of temperature 
(T = 0, 100 MeV and 150 MeV).} 
\label{fig4}
\end{figure}

We study the density dependence of $D$ and $\bar{D}$ masses at 
finite temperatures at selected values of the isospin asymmetric paramater, 
$\eta$ and compare the results with the zero temperature case \cite{amarind}. 
In Fig. \ref{fig4}, we show the variation of the energy of the $D$ mesons 
($D^{+}, D^{0}$) at zero momentum with baryon density $\rho_{B}$ for
different values of isospin asymmetry parameter $\eta$ and with the values 
of temperature as T = 0, 100, 150 MeV. The isospin-asymmetry in the 
medium is seen to give an increase in the  $D^{0}$ mass and a drop in 
the $D^{+}$ mass as compared to the isospin symmetric ($\eta = 0$) case. 
This is observed both for zero temperature \cite{amarind} and finite 
temperature cases. At nuclear matter saturation density, $\rho_{B} = 
\rho_{0}$, the drop  in the mass of $D^{+}$ meson from its vacuum value 
(1869 MeV) is 81 MeV for zero temperature case in isospin symmetric medium. 
At a density of $4\rho_{0}$, this drop in the mass of $D^{+}$ meson is 
seen to be about 364 MeV for $\eta$=0. At finite temperatures, 
the drop in the mass of $D^{+}$ meson for a given value of isospin asymmetry
decreases as compared to the zero temperature case. For example, at 
nuclear saturation density, $\rho_{0}$, the drop 
in the mass of $D^{+}$ meson turns out to be 74, 67 and 63 MeV at a 
temperature of T = 50, 100 and 150 MeV respectively for the isospin symmetric
matter. At a higher density of $\rho_B=4\rho_0$, zero temperature value for
the $D^+$ drop of about 364 MeV is modified to 352, 327 and 307 MeV 
for T = 50, 100 and 150 MeV respectively. Thus the masses of $D$ mesons 
at finite temperatures and finite densities are observed to be larger than
the values at zero temperature case. This is because of the increase in 
the magnitudes of the scalar fields $\sigma$ and $\zeta$ with temperature 
at finite densities as mentioned earlier. The same behaviour remains for 
the isospin asymmetric matter. This behaviour of the nucleons and hence 
of the D-mesons with temperature was also observed earlier for symmetric 
nuclear matter at finite temperatures within the chiral effective model 
\cite{amdmeson}. The drop in the mass of $D^{+}$ meson is seen to be larger 
as we increase the value of the isospin asymmetry parameter. We observe that 
as we change $\eta$ from 0 to 0.5, then the drop in the mass of $D^{+}$ 
meson is 97 MeV and 397 MeV at densities of $\rho_{0}$ and $4\rho_{0}$
respectively, 
for the zero temperature case. At $ T = 50 $ MeV these values change to 91 
MeV at $\rho_{0}$ and 389 MeV at a density of $4\rho_{0}$. At $T = 150$ 
MeV the drop in the mass of $D^{+}$ is 85 MeV at $\rho_{0}$ and 369 MeV at 
$4\rho_{0}$. Thus we observe that for a given value of density, as we move 
from $\eta = 0$ to $\eta = 0.5$ the drop in mass of $D^{+}$ mesons is 
lower at higher temperatures.
 \begin{figure}
\includegraphics[width=16cm,height=16cm]{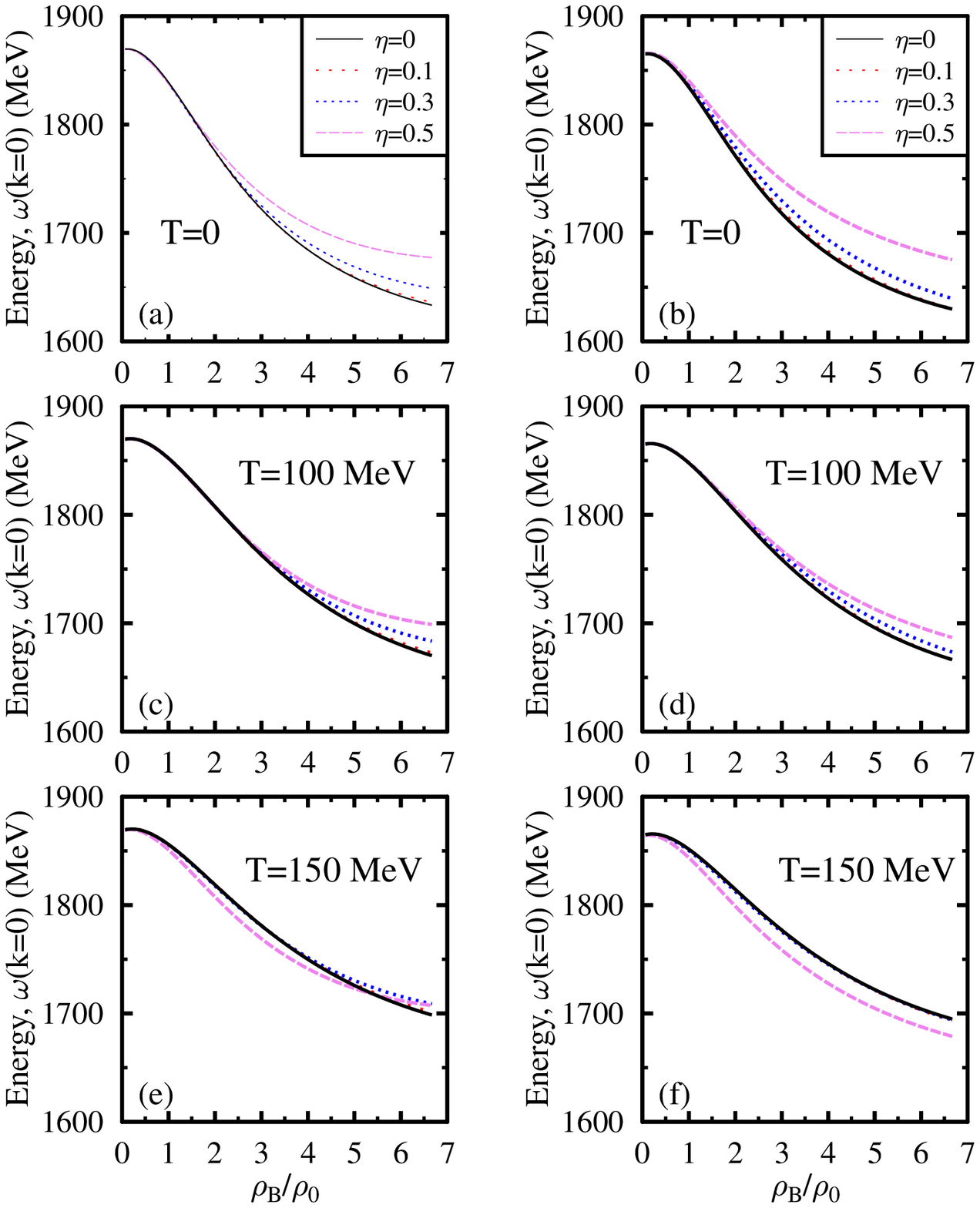} 
\caption{(Color online)  The energies of $D^{-}$ meson ((a),(c) and (e)) 
and of $\bar {D^0}$ meson ((b),(d) and (f)), at momentum $k = 0$, versus 
the baryon density, expressed in units of nuclear saturation density, 
$\rho_{B}/\rho_{0}$, for different values of the isospin asymmetry 
parameter ($\eta = 0, 0.1, 0.3, 0.5$) and for given values of temperature 
(T = 0, 100 MeV and 150 MeV).} 
\label{fig5}
\end{figure}

The mass of the $D^0$ meson drops with density as can be seen from figure 
\ref{fig4}. The drop in the mass of $D^{0}$ meson at density $\rho_{0}$ 
from its vacuum value (1864.5 MeV) is 80, 73.5, 66.5, 63 MeV for temperature 
$T = 0, 50, 100, 150$ MeV 
respectively at $\eta = 0$. At density $4\rho_{0}$ and isospin-asymmetry 
parameter $\eta = 0$, these values become 361, 348.5, 324 and  
304 MeV for temperatures T=0,50,100 and 150 MeV respectively. 
For the $D^{0}$ meson, there is seen to be an increase in the mass 
as we move from isospin symmetric to isospin asymmetric medium. 
For example, at zero temperature and baryon density equal to $\rho_{0}$ 
and $4\rho_{0}$, the rise in the masses of the $D^{0}$ meson are 21 and 91 
MeV respectively as we move from isospin-symmetric medium ($\eta = 0$) 
to isospin-asymmetric medium ($\eta = 0.5$). At a temperature of 50 MeV, 
these values become 19 MeV at $\rho_{0}$ and 85 MeV at $4\rho_{0}$. 
For T=150 MeV, these values become 9 MeV at $\rho_{0}$ and 
43.5 MeV  at $4\rho_{0}$. Thus for $D^{0}$ mesons, the rise in the mass, 
is seen to be lowered at higher temperatures as we move from $\eta$ = 0 to 
$\eta$ = 0.5. 

\begin{figure}
\includegraphics[width=16cm,height=16cm]{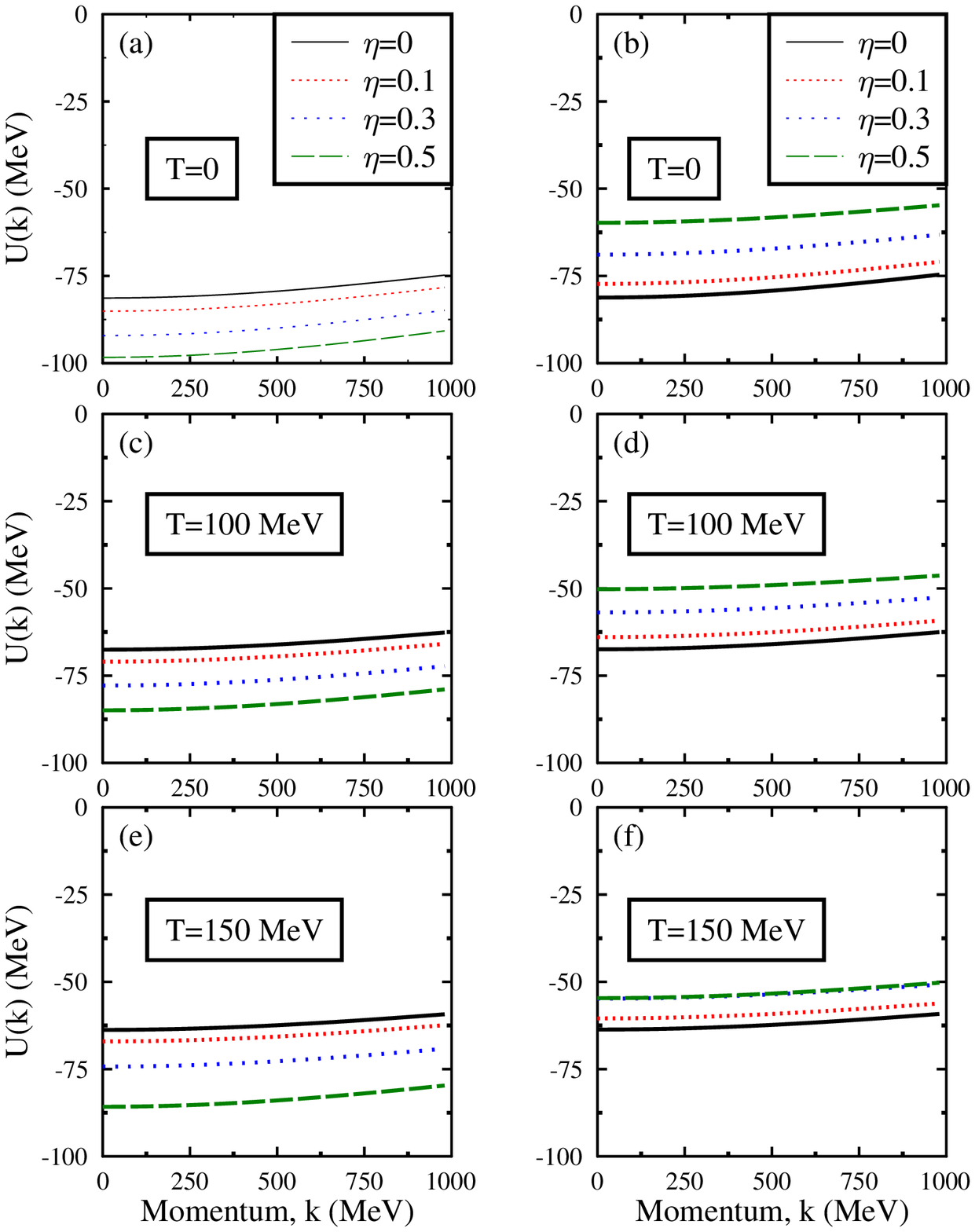} 
\caption{(Color online) The optical potential of $D^{+}$ meson 
(a,c and e) and of $D^{0}$ meson (b,d and f), 
are plotted as functions of momentum for $\rho_{B}=\rho_0$, 
for different values of the isospin asymmetry 
parameter ($\eta = 0, 0.1, 0.3, 0.5$) and for given values 
of temperature (T = 0, 100 MeV and 150 MeV).} 
\label{fig6}
\end{figure}
\begin{figure}
\includegraphics[width=16cm,height=16cm]{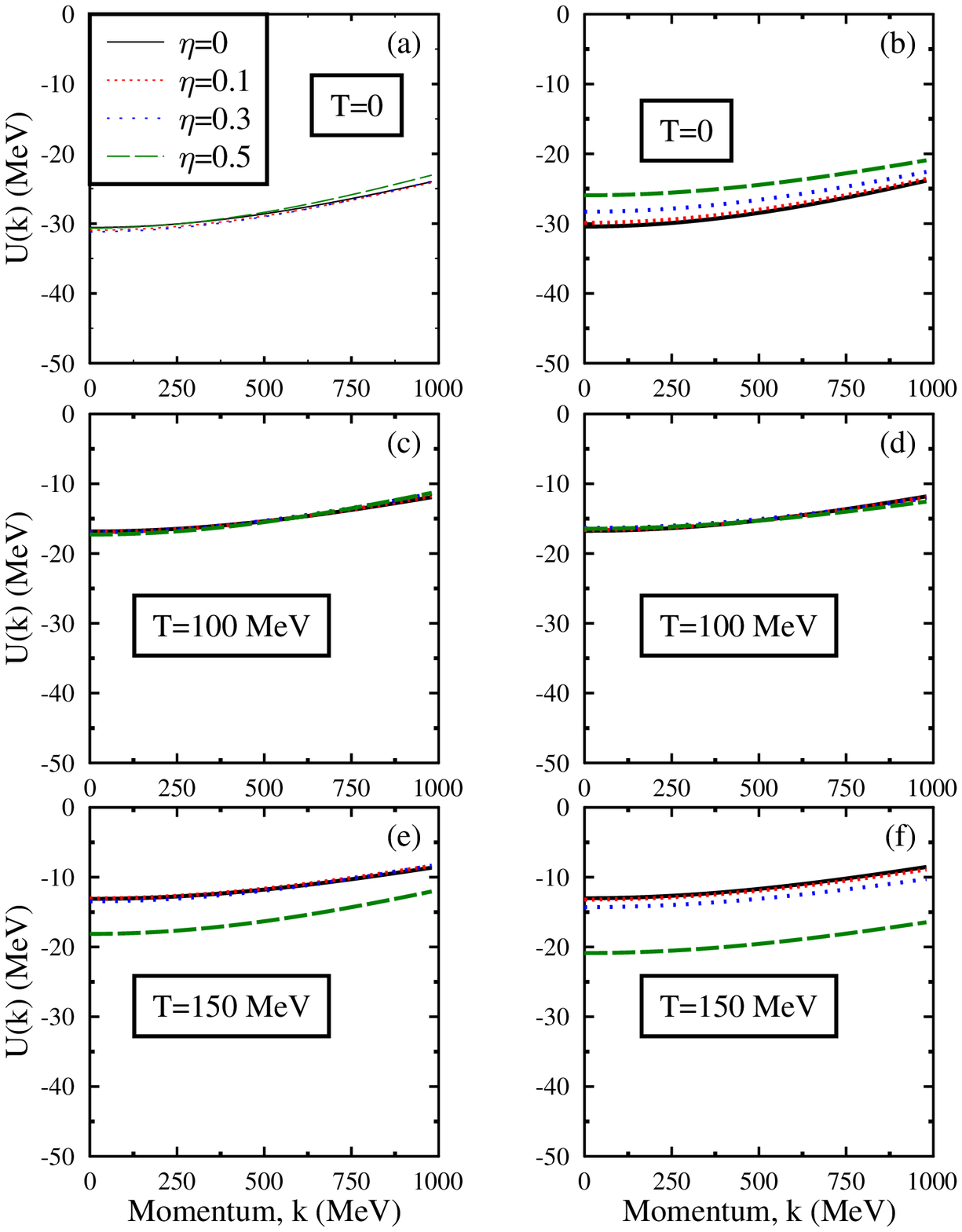} 
\caption{(Color online) The optical potential of $D^{-}$ meson 
(a,c and e) and of $\bar {D^{0}}$ meson (b,d and f), 
are plotted as functions of momentum for $\rho_{B}=\rho_0$, 
for different values of the isospin asymmetry 
parameter ($\eta = 0, 0.1, 0.3, 0.5$) and for given values 
of temperature (T = 0, 100 MeV and 150 MeV).} 
\label{fig7}
\end{figure}
Fig.\ref{fig5} shows the results for the density dependence of the 
energies of the $\bar{D}$ mesons at zero momentum at values of the
temperature, $ T = 0, 100, 150$ MeV. There is seen to be a drop 
of the masses of both the $D^-$ and $\bar {D^0}$ with density.
This is due to the dominance of the attractive scalar exchange 
contribution as well as the range terms (which becomes attractive 
above a density of about 2--2.5 times the nuclear matter saturation 
density) over the repulsive Weinberg-Tomozawa interaction \cite{amarind}. 
It is observed that the drop in the mass of $D^{-}$ ($\bar{D}^{0}$) 
in isospin symmetric nuclear matter is 30 (29.5) MeV at $\rho_{0}$ 
and 184 (183) at $4\rho_{0}$, at zero temperature, from their vacuum 
values. As we go to higher temperatures, the drop in the masses of
$D^{-}$ and $\bar{D}^{0}$ mesons decreases. For example, at $\eta$ = 0 
and $\rho_{B} = \rho_{0}$ the drop in the mass of $D^{-}$ meson is 
23, 16 and 12 MeV at a temperature of 50, 100 and 150 MeV respectively. 
The masses of $D^{-}$ and $\bar{D}^{0}$ mesons are seen to have 
negligible dependence on the isospin asymmetry upto a density of 
$\rho_{B}=\rho_{0}$. However, at high densities there is seen to be 
appreciable dependence of these masses on the parameter, $\eta$. 
For example, at a baryon density of $4\rho_{0}$ and $\eta$ = 0, 
the drop in the mass of the $D^{-}$ meson is 169, 140.5 and 117.5 
MeV for T = 50, 100 and 150 MeV respectively. But at  $\eta$ = 0.5,
these values change to 150, 131 and 126 MeV for T = 50, 100 and 150 MeV 
respectively. It is seen that, at high densities there is an increase 
in the masses of both $D^{-}$ and $\bar{D}^{0}$ mesons in isospin 
asymmetric medium as compared to those in the isospin symmetric 
nuclear matter for temperatures T=0, 50 and 100 MeV. However, 
at $T = 150$ MeV, it is observed that for densities upto about 
5$\rho_0$, the mass of $D^{-}$ meson is higher in the isospin 
symmetric matter as compared to in the isospin asymmetric matter 
with $\eta$ = 0.5. It is also seen that the modifications in the 
masses of $D^{-}$ mesons is negligible as we change $\eta$ from
0 to 0.3 upto a density of about 4$\rho_0$. For the $\bar{D}^{0}$ 
meson, one sees that the isospin dependence is negligible upto 
$\eta$=0.3 and the mass is lowered for $\eta$=0.5 as compared to 
the mass in symmetric nuclear matter. The drop in the mass of 
$\bar{D}^{0}$ mesons is 8 MeV and 18 MeV at $\rho_{0}$ and 4$\rho_{0}$ 
respectively. For $\bar{D}^{0}$ mesons also there is seen to be negligible 
change in mass as we change the aysmmetry paramater $\eta$ from 0 upto 0.3. 
This is because the drop in the mass 
of $\bar{D}^{0}$ mesons given by Weinberg-Tomozawa term almost cancel with 
the increase due to the scalar and range terms as we go from $ \eta = 0$ to 
$ \eta = 0.3$. At zero temperature \cite{amarind} as well as for 
temperatures T=50 and 100 MeV, there is increase in the mass of the 
$\bar{D}$ mesons ($D^-$,$\bar {D^0}$) as we go from isospin symmetric 
medium to the isospin asymmetric medium. This is because 
for T=0,50 and 100 MeV, the increase in mass of 
$\bar D$  given by the scalar exchange and the range terms were 
dominating over the drop given by the Weinberg Tomozawa term as 
we go from symmetric nuclear medium ($\eta = 0$) to isospin asymmetric 
nuclear medium ($\eta$ =0.1,0.3 0.5). However, at $T = 150$ MeV,
for $\eta$ =0.5, the drop given by Weinberg term dominates over 
the rise given by scalar and range terms for $\bar {D^0}$ and upto
a density of about 5$\rho_0$ for $D^-$, and therefore mass of 
$\bar{D}$ mesons decreases as we go from symmetric nuclear medium 
to isospin asymmetric nuclear medium in these density regimes.

\begin{figure}
\includegraphics[width=16cm,height=16cm]{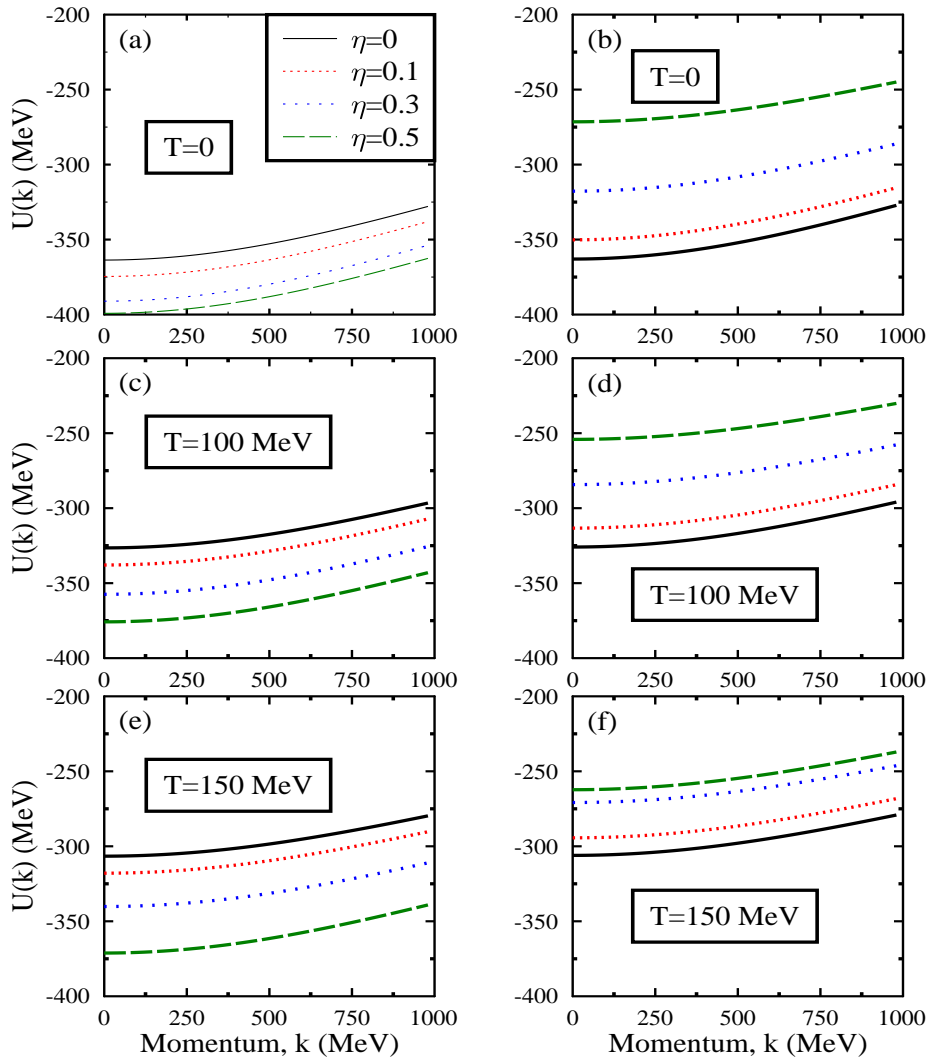} 
\caption{(Color online) The optical potential of $D^{+}$ meson 
(a,c and e) and of $D^{0}$ meson (b,d and f), 
are plotted as functions of momentum for $\rho_{B}=4\rho_0$, 
for different values of the isospin asymmetry 
parameter ($\eta = 0, 0.1, 0.3, 0.5$) and for given values 
of temperature (T = 0, 100 MeV and 150 MeV).} 
\label{fig8}
\end{figure}

\begin{figure}
\includegraphics[width=16cm,height=16cm]{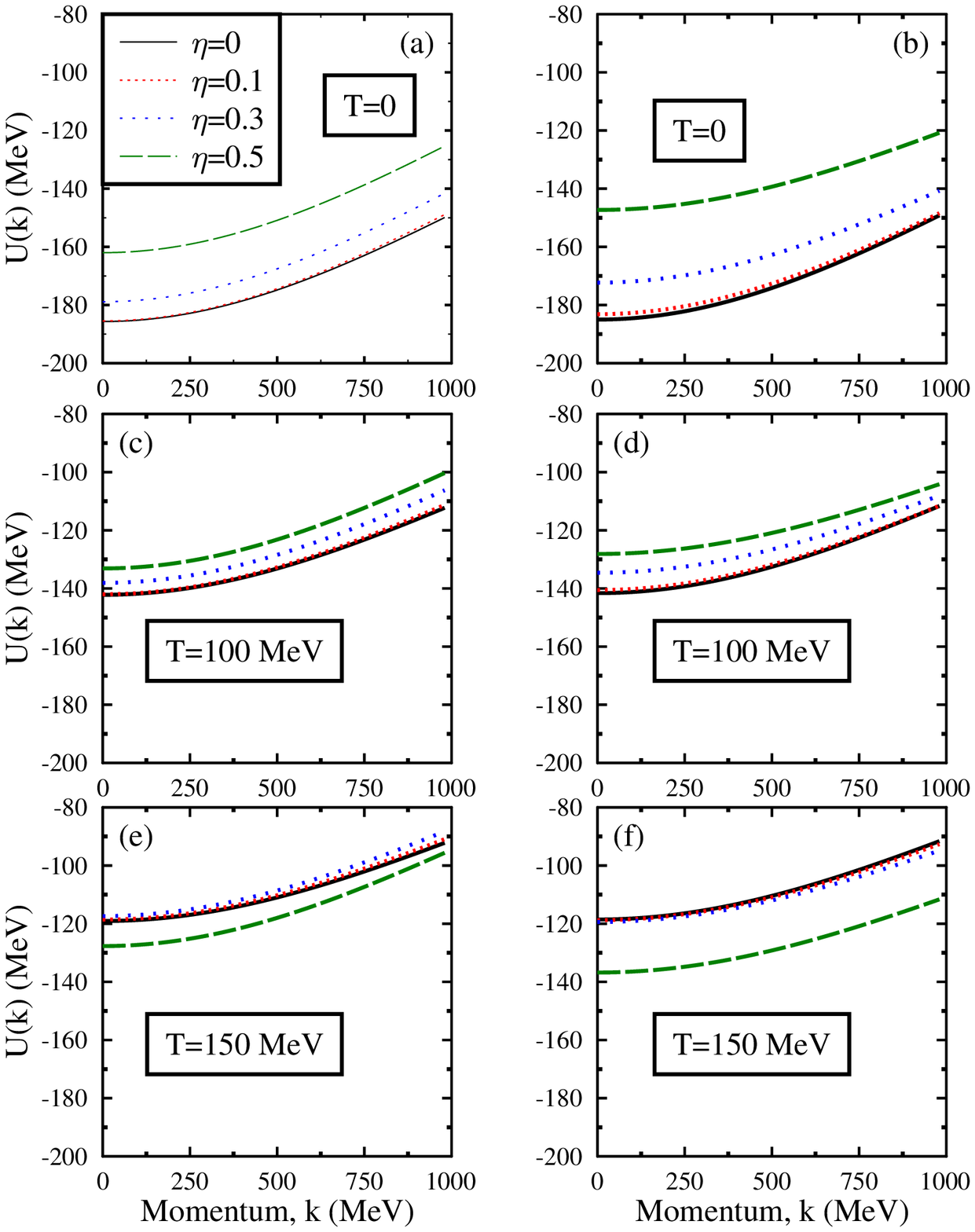} 
\caption{(Color online) The optical potential of $D^{-}$ meson 
(a,c and e) and of $\bar {D^{0}}$ meson (b,d and f), 
are plotted as functions of momentum for $\rho_{B}=4\rho_0$, 
for different values of the isospin asymmetry 
parameter ($\eta = 0, 0.1, 0.3, 0.5$) and for given values 
of temperature (T = 0, 100 MeV and 150 MeV).} 
\label{fig9}
\end{figure}

Figures \ref{fig6} and \ref{fig8} show the isospin dependence
of the optical potentials for the $D$ mesons as functions of the momentum, 
for densities $\rho_{0}$ and $4\rho_{0}$ respectively and for values of
the temperature as $T = 0, 100, 150$ MeV. Figures \ref{fig7} 
and \ref{fig9} illustrate the optical potentials for the 
$\bar{D}$ doublet. The isospin dependence of optical potentials 
is seen to be quite significant for high densities for the D-meson doublet 
($D^{+}, D^{0}$) as compared to those for the $\bar D$ doublet.
This is a reflection of the strong isospin dependence of the
masses of the D-mesons as compared to the $\bar {\rm D}$ as has been
already illustrated in figures \ref{fig4} and \ref{fig5}.
For the $\bar{D}$ mesons, it is seen, from figure \ref{fig5}, 
that the masses of the the $D^{-}$ meson and $\bar{D}^{0}$ meson 
for a fixed value of the isospin asymmetry parameter, $\eta$ are 
very similar, an observation which was seen earlier for the
zero temperature case \cite{amarind}. These are reflected in
their optical potentials, plotted in figures \ref{fig7}
and \ref{fig9}, where one sees a maximum difference of 
about 5 MeV or so between $D^-$ and $\bar {D^0}$ for $\rho_B=\rho_0$
and about 10 -- 15 MeV for $\rho_B=4\rho_0$. 
The present investigations of the optical potentials for the
$D$ and $\bar D$ mesons show a much stronger dependence of
isospin asymmetry on the $D$ meson doublet, as compared to
that in the $\bar D$ meson doublet, as was already observed for
the zero temperature case. 
The medium modifications of the masses of $D$ and $\bar {D}$ mesons can 
lead to the explanation of $J/\psi$ suppression  observed by NA50 
collaboration at $158$ GeV/nucleon in the Pb-Pb collisions \cite{blaiz}. 
Due to the drop in the mass of the $D\bar D$ pair in the nuclear medium,
it can become a possibility that the excited states of charmonium 
($\psi^{'}, \chi_{c2}, \chi_{c1}, \chi_{c0}$) can decay to $D\bar{D}$
\cite{amarind} and hence the production of $J/\Psi$ from the decay of 
these excited states can be suppressed. Even at high values of densities 
at given temperatures, it can become a possibility that $J/\psi$ itself
decays to $D\bar{D}$ pairs. Thus the medium modifications of the $D$ mesons 
can change the decay widths of the charmonium states \cite{friman}. 

The decay of the charmonium states have been studied in Ref. 
\cite{friman,brat6}. It is seen to depend sensitively on the relative 
momentum in the final state. These excited states might become narrow 
\cite{friman} though the $D$ meson mass is decreased appreciably at 
high densities. It may even vanish at certain momentum corresponding 
to nodes in the wave function \cite{friman}. Though the decay widths 
for these excited states can be modified by their wave functions, the 
partial decay width of $\chi_{c2}$, owing to absence of any nodes, 
can increase monotonically with the drop of the $D^{+}D^{-}$ pair mass 
in the medium. This can give rise to depletion in the  $J/\Psi$ yield 
in heavy-ion collisions. The dissociation of the quarkonium states 
($\Psi^{'}$,$\chi_{c}$, $J/\Psi$) into $D\bar{D}$ pairs has also been 
studied \cite{wong,digal} by comparing their binding energies with the 
lattice results on the temperature dependence of the heavy-quark 
effective potential \cite{lattice}.

\section{summary}
We have investigated in a chiral model the in-medium masses of the $D$, 
$\bar{D}$ mesons in hot isospin asymmetric nuclear matter, arising from 
their interactions with the nucleons and the scalar mesons. The properties 
of the light hadrons -- as studied in  $SU(3)$ chiral model -- modify the 
$D(\bar{D})$ meson properties in the dense and hot hadronic matter. The 
$SU(3)$ model, with parameters fixed from the properties of the hadron 
masses in vacuum and low-energy KN scattering data, is extended to 
SU(4) to drive the interactions of $D(\bar{D})$ mesons with the light 
hadron sector. The mass modifications of $D^{+}$ and $D^{0}$ mesons is 
strongly dependent on isospin-asymmetry of medium. The mass of $D$ mesons 
observed at finite temperature is more as compared to zero temperature 
case because of less decrease of scalar 
fields at finite temperature and finite densities. This is in accordance 
with earlier work on $D$ mesons modifications in symmetric nuclear matter 
at finite temperature \cite{amdmeson}. The mass modification for the $D$ 
mesons are seen to be similar to earlier finite density calcualtions of 
QCD sum rules \cite{qcdsum08,weise} as well as to the quark-meson coupling 
model \cite{qmc}, in contrast to the small mass modificatins in the coupled 
channel approach \cite{ljhs,mizutani8}. Also we obtained small attractive 
mass shifts for $\bar{D}$ mesons in contrast with coupled channel approach 
\cite{mizutani8}. These attractive potentials for $\bar{D}$ mesons are in 
favor of charmed mesic nuclei. In our calculations the presence of the 
repulsive  first range term (with coefficient $-\frac{1}{f_{D}}$ in 
Eq. (\ref{ldn})) is compensated by the attractive $d_{1}$ and $d_{2}$ 
terms  in Eq.(\ref{ldn}). Among attractive $d_{1}$ and $d_{2}$ terms, 
 $d_{1}$ term is found to be dominating over $d_{2}$ term. 

The medium modifications of the $D$ meson masses can lead to a suppression 
in the $J/\Psi$ yield in heavy-ion collisions, since the excited states of 
the $J/\Psi$ ($\simeq 5\rho_{0}$), $J/\Psi$, can decay to $D\bar{D}$ pairs 
in the dense hadronic medium. The decay to the $D^{+}D^{-}$ pairs seems 
to be insensitive at zero temperature but at high temperture like $150$ MeV 
these decay become sensitive to isospin asymmetry of the medium. The 
isospin asymmetry lowers the density at which decay to $D^{+}D^{-}$ 
pairs occur. Due to increase in the mass of $D^{0}\bar{D}^{0}$ in the 
isospin-asymmetric medium, isospin-asymmetry is seen to disfavor the 
decay of the charmonium states to the $D^{0}\bar{D}^{0}$  pairs. At a
finite temperature there does not seem to be the possibility of decay 
of $J/\Psi$ to  $D^{0}\bar{D}^{0}$  pairs. The isospin dependence of 
$D^{+}$  and $D^{0}$ masses is seen to be a dominant medium effect 
at high densities, which might show in their production ($D^{+}/D^{0}$), 
whereas, for the $D^{-}$ and $\bar{D}^{0}$, one sees that, even though 
these have a strong density dependence, their in-medium masses remain 
similar at a given value for the isospin-asymmetry parameter $\eta$. 
The strong density dependence as well as the isospin dependence of 
the $D(\bar{D})$ meson optical potentials in asymmetric nuclear matter 
can be tested in the asymmetric heavy-ion collision experiments at 
future GSI facility \cite{gsi}. In this work we considered only nucleons. 
The study of the in-medium modifications of $D$ mesons in hyperonic 
matter along with nucleons at zero and finite temperatures within 
the chiral $SU(4)$ model, and the study of in-medium properties
of the charmonia states in the hot asymmetric hadronic matter,
due to their interactions with the D-mesons as well as interactions
to the dilaton field associated with the broken scale invariance
(related to the gluon condensates) will be possible extensions of 
the present investigation.

\acknowledgements
Financial support from Department of Science and Technology, Government 
of India (project no. SR/S2/HEP-21/2006) is gratefully acknowledged 
by the authors. One of the authors (AM) is grateful to the Frankfurt
Institute of Advanced Studies (FIAS), University of frankfurt, 
for warm hospitality and acknowledges financial support from 
Alexander von Humboldt Stiftung when this work was initiated.


\begin{thebibliography}{1}
\bibitem {ceres}
	G. Agakichiev {\it et al.} (CERES collaboration),
		Phys. Rev. Lett. {\bf 75}, 1272 (1995);
	G. Agakichiev {\it et al.} (CERES collaboration),
		Phys. Lett. B {\bf 422}, 405 (1998);
	G. Agakichiev {\it et al.} (CERES collaboration), Nucl. Phys.
		{\bf A661}, 23c (1999).
\bibitem{helios}
	N. Masera {\it et al.} (HELIOS-3 collaboration),
	Nucl. Phys. {\bf A590}, 93c (1995).
\bibitem{dls} 
 R. J. Porter {\it et al.} (DLS collaboration),
 Phys. Rev. Lett. {\bf 79}, 1229 (1997); 
 W. K. Wilson  {\it et al.} (DLS collaboration),
  Phys. Rev. C {\bf 57}, 1865 (1998).
\bibitem{Brat1}
	E. L. Bratkovskaya and W. Cassing, Nucl. Phys.
	{\bf A619}, 413 (1997).
\bibitem{CB99}
	W. Cassing and E. L. Bratkovskaya, Phys. Rep. {\bf 308}, 65 (1999).
\bibitem{vecmass}
	A. Mishra, J. C. Parikh and W. Greiner, J. Phys. G {\bf 28},
	151 (2002).
\bibitem{dilepton}
	A. Mishra, J. Reinhardt, H. St\"ocker and W. Greiner,
	Phys. Rev. C {\bf 66}, 064902 (2002).
\bibitem{liko}
	G. Q. Li, C. M. Ko and G. E. Brown,
	Nucl. Phys. {\bf A606}, 568 (1996).
\bibitem{cmko}
	C. M. Ko, J. Phys. G {\bf 27}, 327 (2001).
\bibitem{lix}
	G. Q. Li, C.-H. Lee and G. E. Brown,
	Nucl. Phys. {\bf A625}, 372 (1997).
\bibitem{Li2001}
	S. Pal, C. M. Ko, and Z.-W. Lin, Phys. Rev. C {\bf 64}, 042201
  	(2001).
\bibitem{K5}
	F. Laue {\it et al.}, Phys. Rev. Lett. {\bf 82}, 1640 (1999).
\bibitem{K6}
	C. Sturm {\it et al.}, Phys. Rev. Lett. {\bf 86}, 39 (2001).
\bibitem{K4}
	A.  F\"orster {\it et al.}, KaoS Collaboration,
	J. Phys. G {\bf 28}, 2011 (2002).
\bibitem{kaosnew}
	M. Menzel {\it et al.}, Phys. Lett. B {\bf 495}, 26 (2000).
\bibitem{gsi}
	see e.g. http://www.gsi.de/fair/experiments/CBM/
\bibitem{NA501}
	M. Gonin {\it et al.}, NA50 Collaboration,
	Nucl. Phys. {\bf A610}, 404c (1996).
\bibitem{NA50e}
	M. C. Abreu {\it et al.}, NA50 Collaboration,
	Eur. Phys. J. C {\bf 14}, 443 (200).
\bibitem{NA502}
	L. Rammello {\it et al.}, NA50 Collaboration,
	Nucl. Phys. {\bf A638}, 261c (1998);
	M. C. Abreu {\it et al.}, Phys. Lett. B {450}, 456 (1999);
       L. Ramello {\it et al.}, Nucl. Phys. {\bf A715}, 242 (2003).
\bibitem{blaiz}
	J. P. Blaizot and J. Y. Ollitrault, Phys. Rev. Lett.
	{\bf 77}, 1703 (1996).       	
\bibitem{satz}
	T. Matsui and H. Satz, Phys. Lett. B {\bf 178}, 416 (1986).  
\bibitem{zhang}
	B. Zhang, C. M. Ko, B. A. Li, Z. Lin, and B. H. Sa,
	Phys. Rev. C {\bf 62}, 054905 (2000).
\bibitem{brat5}
	W. Cassing and E. L. Bratkovskaya,
	Nucl. Phys. {\bf A623}, 570 (1997).
\bibitem{elena}
	E. L. Bratkovskaya, W. Cassing and H. St\"ocker,
	Phys. Rev. C {\bf 67}, 054905 (2003).
\bibitem{brat6}
	Ye. S. Golubeva, E. L. Bratkovskaya, W. Cassing,
	and L. A. Kondratyuk, Eur. Phys. J. A {\bf 17}, 275 (2003).
\bibitem {pAdata}
	L. Antoniazzi {\it et al.}, E705 Collaboration,
	Phys. Rev. Lett. {\bf 70}, 383 (1993);
	Y. Lemoigne {\it et al.},	Phys. Lett. B {\bf 113}, 509 (1982).
\bibitem{leeko}
        S. H. Lee and C. M. Ko, Phys. Rev. C {\bf 67}, 038202 (2003).
\bibitem{haya1}
	A. Hayashigaki, arXiv:9811092v1 [nucl-th].
\bibitem{friman}
	B. Friman, S. H. Lee and T. Song,
	Phys. Lett. B {\bf 548}, 153 (2002).	        			
\bibitem{arata}
	A. Hayashigaki, Phys. Lett. B {\bf 487}, 96 (2000).
\bibitem{qcdsum08} 
        T. Higler, R. Thomas, B. K\"ampfer, nucl-th/0809.4996.
\bibitem{qmc}
	K. Tsushima, D. H. Lu, A. W. Thomas, K. Saito, and R. H. Landau,
	Phys. Rev. C {\bf 59}, 2824 (1999);
	A. Sibirtsev,	K. Tsushima, and A. W. Thomas,
	Eur. Phys. J. A {\bf 6}, 351 (1999).
\bibitem{paper3}
 	P. Papazoglou, D. Zschiesche, S. Schramm, J. Schaffner-Bielich,
	H. St\"ocker, and W. Greiner, Phys. Rev. C {\bf 59},  411  (1999).
\bibitem{kmeson}
A. Mishra, E. L. Bratkovskaya, J. Schaffner-Bielich, S. Schramm
     and H. St\"ocker, Phys. Rev. C {\bf 70}, 044904 (2004).
\bibitem{isoamss}
A. Mishra and S. Schramm, Phys. Rev. C {\bf 74}, 064904 (2006),	
A. Mishra, S. Schramm and W. Greiner, Phys. Rev. C {\bf 78}, 024901 (2008).	
\bibitem{isoamss2}
Amruta Mishra, Arvind Kumar, Sambuddha Sanyal, S. Schramm,
Eur. Phys, J. A {\bf 41}, 205  (2009).
\bibitem{weinberg}
S.Weinberg, Phys. Rev. {\bf 166} 1568 (1968).
\bibitem{coleman}
S. Coleman, J. Wess, B. Zumino, Phys. Rev. {\bf 177} 2239 (1969);
C.G. Callan, S. Coleman, J. Wess, B. Zumino, Phys. Rev. {\bf 177}
2247 (1969).
\bibitem{bardeen}
W. A. Bardeen and B. W. Lee, Phys. Rev. {\bf 177} 2389 (1969).
\bibitem{hartree}
	D. Zschiesche, A. Mishra, S. Schramm, H. St\"ocker and W. Greiner,
        Phys. Rev. C {\bf 70}, 045202 (2004).
\bibitem{kristof1}
	A. Mishra, K. Balazs, D. Zschiesche, S. Schramm,
	H. St\"ocker, and W. Greiner,
        Phys. Rev. C {\bf 69}, 024903 (2004).
\bibitem {amdmeson} 
A. Mishra, E. L. Bratkovskaya, J. Schaffner-Bielich, 
S.Schramm and H. St\"ocker, Phys. Rev. {\bf C 69}, 015202 (2004). 
\bibitem{amarind} 
Amruta Mishra and Arindam Mazumdar, 
Phys. Rev. C {\bf 79},  024908 (2009).
\bibitem{ltolos} L.Tolos, J. Schaffner-Bielich and A. Mishra,
Phys. Rev. {\bf C 70}, 025203 (2004).
\bibitem{kbarn}
E.A. Veit, B. K. Jennings, A. W. Thomas, R. C. Berrett, Phys. Rev.
{\bf D 31}, 1033 (1985);
L. Tolos, A. Ramos, A. Polls and T.T.S. Kuo, Nucl. Phys. {\bf A 690}, 
547 (2001);
V. Koch, Phys. Lett. {\bf B 337}, 7 (1994); P. B. Siegel and W. Weise,
Phys. Rev. {\bf C 38}, 2221 (1998).
\bibitem{ljhs}
L. Tolos, J. Schaffner-Bielich
and H. St\"ocker, Phys. Lett. {\bf B 635}, 85 (2006).
\bibitem{HL} J. Hofmann and M.F.M.Lutz, Nucl. Phys. {\bf A 763}, 90 (2005).
\bibitem{KSFR} K. Kawarabayashi and M. Suzuki, Phys. Rev. Lett. {\bf 16},
255 (1966); Riazuddin and Fayyazuddin, Phys. Rev. {\bf 147}, 1071 (1966).
\bibitem{mizutani6}
 T. Mizutani and A. Ramos, Phys. Rev. {\bf C 74},
065201 (2006).
\bibitem{mizutani8} 
L.Tolos, A. Ramos and T. Mizutani, Phys. Rev. {\bf C 77},
015207 (2008).
\bibitem{MK} M.F.M. Lutz and C.L. Korpa, Phys. Lett. {\bf B 633}, 43 (2006).
\bibitem{tolosra} 
Laura Tolos, Raquel Molina, Daniel Gamermann and Eulogio Oset, arXiv:0901.1588v1 [nucl-th]. 
\bibitem {borasoy}
B.Borasoy and U-G. Meissner, Int. J. Mod. Phys. A {\bf 11} 5183 (1996).
\bibitem{wong} C. Y. Wong T. Barnes, E. S. Swanson and H. W. Crater,
nucl-th/0112023.
\bibitem{roeder}
	D. R\"oder, J. Ruppert, and D.H. Rischke,
	Phys. Rev. D {\bf 68}, 016003 (2003).
\bibitem{digal}
	S. Digal, P. Petreczky and H. Satz,
	Phys. Lett. B {\bf 514}, 57 (2001).
\bibitem{lattice}
	F. Karsch, E. Laermann and A. Peikert, Nucl. Phys. B {\bf 605}, 579 (2001);
    F. Karsch {\it et al.}, Phys. Lett. B {\bf 502}, 321 (2001).
\bibitem{weise}
	P. Morath, W. Weise and S. H. Lee, 17th Autumn school:
	Lisbon 1999, QCD: { Perturbative or Nonperturbative}?
	Singapore: World Scientific, p. 425.    			             
\end{thebibliography}
\end{document}